\documentclass[12pt]{iopart}
\usepackage{graphicx}
\usepackage{iopams}

\begin{document}

\title[Spin current in an electron waveguide tunnel-coupled to topological insulator]{Spin current in an electron waveguide tunnel-coupled to topological insulator}

\author{Aleksei A Sukhanov and Vladimir A Sablikov}

\address{V A Kotel’nikov Institute of Radio Engineering and Electronics, Russian Academy of Sciences,
Fryazino, Moscow District, 141190, Russia}
\ead{AASukhanov@yandex.ru}
\begin{abstract}
We show that electron tunneling from edge states in two-dimensional topological insulator into a parallel electron waveguide leads to the appearance of spin-polarized current in the waveguide. The spin polarization $P$ can be very close to unity and the electron current passing through the tunnel contact splits in the waveguide into two branches flowing from the contact. The polarization essentially depends on the electron scattering by the contact and the electron-electron interaction in the one-dimensional edge states. The electron-electron interaction is treated within the Luttinger liquid model. The main effect of the interaction stems from the renormalization of the electron velocity, due to which the polarization increases with the interaction strength. Electron scattering by the contact leads to a decrease in $P$. A specific effect occurs when the bottom of the subbands in the waveguide crosses the Dirac point of the spectrum of edge states when changing the voltage or chemical potential. This leads to changing the direction of the spin current.
\end{abstract}

\pacs{72.25.Dc, 73.40.Gk, 73.63.Nm}

\maketitle

\section{Introduction}

Spin properties of surface and edge states are hallmark of topological insulators (TIs)~\cite{Hasan,Qi} which attract much interest as a playground for fundamental physics and a basis for potential applications in spintronics and quantum computing. The surface states in three-dimensional TIs and edge states in two-dimensional (2D) TIs have a helically polarized spin structure. The spin of electrons is strongly coupled to their momentum. A consequence of this coupling is that the edge states carry a spin current. This fact stimulates the idea of creating spin currents in a nonmagnetic conductor, which is in contact with a TI. Indeed, if the tunneling electrons conserve their spin and direction of the velocity, the spin filtering of electrons occurs in the nonmagnetic conductor since the electrons with different spins move with different velocities.

The existence of the helical edge states with Dirac dispersion in 2D TIs was proved by recent experiments~\cite{Konig,Roth,Brune,Gusev}. Rapid progress in nanotechnology allows one to expect the creation of mesoscopic heterostructures containing TIs~\cite{Brune,Kong}. Therefore it is interesting to consider a model system in which a 2D TI is coupled to a strip of normal 2D electron gas via a tunnel contact. Tunneling of electrons from the edge states in 2D TI into the normal conductor results in the transition of electrons with different spins in the states with different momenta. In that way a spin current arises in the normal conductor even if the barrier is nonmagnetic and external magnetic field is absent.

The present paper aims to study the spin polarization of the tunnel current in the normal conductor, to clarify what factors affect the polarization and estimate the polarization for realistic conditions. The edge states are one-dimensional (1D) and therefore electron-electron interaction plays an essential role in their tunneling. The 1D electrons in these states can be considered as a helical Luttinger liquid~\cite{Wu}. The interaction is known to change the one-particle density of states and to renormalize the electron velocity~\cite{Gimarchi}. Both these factors affect the tunnel current. The tunneling between parallel 1D systems is a subject of high activity in last decade. Semiconductor heterostructures with tunnel-coupled quantum wires were created, and impressive magneto-tunneling experiments were carried out~\cite{Auslander}, which have revealed many-particle effects. The theoretical works, which are closest to the present paper, have studied the electron-electron interaction effects on the tunnel current~\cite{Governale1}, the dependence of the current on the contact length~\cite{Boese}, and such many-particle effects as fractional quasiparticles  and spin-charge separation~\cite{Carpentier}. Spin transport has been investigated much less. Spin filtering of the tunnel current was explored in the case where the spin-orbit interaction acts in one of parallel quantum wires~\cite{Governale2}. Recently the theory of tunneling between 1D systems was extended to the helical Luttinger liquids~\cite{Dolcetto}. Spin injection from helical edge states into a normal electron gas was not studied.

We solve this problem by considering a model system consisting of a 2D TI coupled to a strip of normal 2D electron gas via a tunnel contact of finite length. Electrons in the edge states are treated as helical Luttinger liquid and the strip is considered as a multimode electron waveguide. We have found that the electron current flowing through the tunnel contact into the normal conductor is split there into two currents of electrons with opposite spins flowing in opposite directions. Thus, the electron current in each branch of the waveguide outside the contact equals half of the total current through the tunnel contact. It is interesting that the electron current in the normal conductor is highly spin polarized and there is a spin current. The spin polarization $P$ can be very close to unity. The polarization essentially depends on the interelectron interaction in the Luttinger liquid and the electron scattering by the contact. The main effect of electron-electron interaction is caused by the renormalization of the electron velocity in the edge states of the TI. Increasing the interaction results in the growth of the polarization. The scattering by the contact leads to a decrease of $P$. A specific effect arises when the applied voltage is varied so that a subband bottom in the waveguide crosses the Dirac point of the edge state spectrum. Under this condition the sign of the spin current is changed.

The paper is organized as follows. Section~\ref{model} describes the model we use for calculations and the qualitative discussion of the main issues. The formalism of calculations is presented in Sec.~\ref{formalism}. Main results are presented and discussed in Sec.~\ref{results}. Sec.~\ref{conclude} contains concluding remarks.

\section{The model}\label{model}
\begin{figure}
 \centerline{\includegraphics[width=10cm]{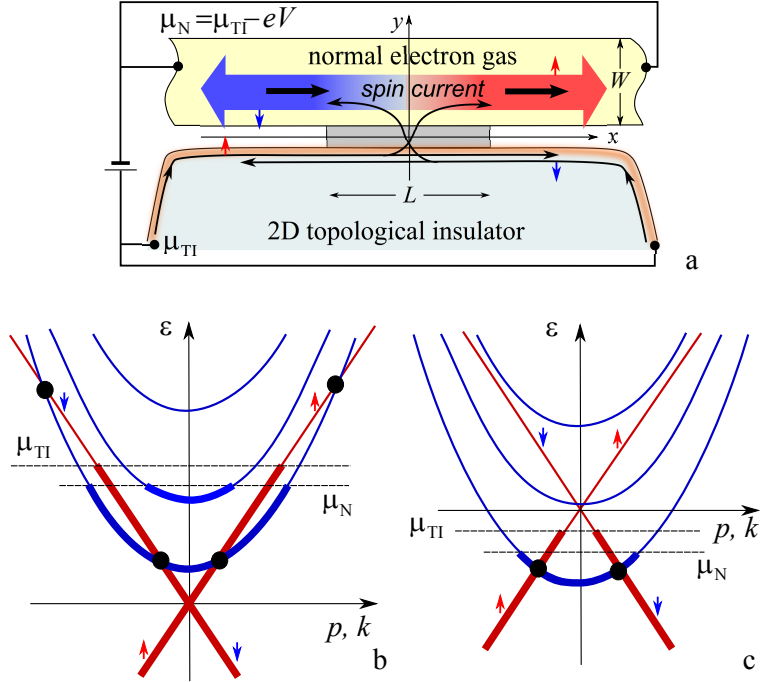}}
 \caption{(Color online) (a) Tunnel contact of 2D TI and the strip of 2D normal electron gas. (b) Energy spectra of edge states (straight lines) in the TI and electrons in the strip (parabolas) in the case where the first subband is above the Dirac point and $\mu_{T\!I}>\mu_N>0$. Occupied states are shown by heavy lines. (c) Energy spectra in the case where the first subband is below the Dirac point and $\mu_N<\mu_{T\!I}<0$.}
\label{TI-N_contact}
\end{figure}

To study the spin-polarized current, which can be created due to the tunneling of electrons from a 2D TI, we consider the model system shown in Fig.~\ref{TI-N_contact}. The TI occupies half-plane $y<0$. At the line $y=0$, it is coupled to a 2D strip of normal electron gas via the tunnel barrier of length $L$. A voltage $V$ is applied between the TI and the strip so that the chemical potential of the edge states $\mu_{T\!I}$ is higher than the chemical potential in the strip $\mu_N$. Electrons in edge states have gapless Dirac-like spectrum, with the spin being strongly coupled to the momentum $p$: the spin of the right-moving electrons is directed upwards and the electrons moving to the left have a spin down. In the strip of width $W$, the electron spectrum is quantized in transverse direction. Within each subband the spectrum is parabolic in the momentum $k$. The choice of a normal electron gas, i.e. an electron gas without spin-orbit interaction, as a medium into which electrons tunnel from the TI, is caused by the fact that here the spin current is uniquely defined and can be measured experimentally (see a discussion in Ref.~\cite{SST} and references there).

Spin current depends on the energy of tunneling electrons. Qualitatively the tunneling process looks as follows. At the beginning let us neglect the scattering by the contact. If the energy is larger than the Dirac point of the edge state spectrum, right-moving electrons with spin up pass from the TI into right-moving states in the strip. In contrast, the left-moving electrons with spin down come to the left-moving states. Thus the tunnel current is split in the strip into two opposite electron flows with opposite spins, as it is shown in Fig.~\ref{TI-N_contact}a,b. In that way the spin current arises.

If the energy is below the Dirac point, the picture is changed as shown in Fig.~\ref{TI-N_contact}c. The right-moving electrons in the TI tunnel into the left-moving states in the strip and vise-versa, the left-moving electrons in the TI come to the right-moving states. Correspondingly, in this case the spin current has opposite direction.

Thus, the shift of the chemical potentials of the TI and the normal electron gas, $\mu_{T\!I}$ and $\mu_N$, through the Dirac point results in the change of the sign of the spin current.

In a realistic situation, there are other factors, which strongly affect the polarization in addition to the conservation of the spin and velocity direction discussed above. These are the scattering of electrons by the contact, due to which some electrons can change the direction of their velocity~\cite{Governale1,Dolcetto}, and the electron-electron interaction, since it substantially changes the spectral function of electrons in 1D edge states and renormalizes the electron velocity~\cite{Gimarchi,Dolcetto}.

\section{Formalism}\label{formalism}

The total Hamiltonian of the system is
\begin{equation}
H=H_{ES}+H_N+H_{tunn}\,,
\end{equation} 
where $H_{ES}$ is the Hamiltonian of the isolated edge states, $H_N$
the Hamiltonian of the isolated strip, and $H_{tunn}$ describes the tunneling between the edge states and the strip.

Electrons in the edge states can be considered as a helical Luttinger liquid~\cite{Wu} and described by bosonized Hamiltonian~\cite{Voit}
\begin{equation}
 H_{ES}=\frac{v_F}{2}\int_{-\infty}^{\infty}dx\left[\Pi^2(x)+\frac{1}{g^2}\left(\frac{\partial
\theta(x)}{ \partial x}\right)^2 \right]\,,
\end{equation} 
where $\theta(x)$ and $\Pi(x)$ are conjugated bosonic phases, $v_F$ the Fermi velocity, and $g$ the interaction parameter. The 1D fermion creation operator is presented in terms of chiral fields $\psi^{\dag}_{ES,r}$ for right and left-moving fermions ($r=\pm1$)
\begin{equation}
 \psi^{\dag}_{ES}=\sum_{r=\pm}\psi^{\dag}_{ES,r},
\end{equation} 
with $\psi^{\dag}_{ES,r}$ being
\begin{equation}
 \psi^{\dag}_{ES,r}=\sqrt{\frac{\omega_c}{2\pi v}}e^{irp_Fx}\exp\left\{i\sqrt{\pi}\left[
r\theta(x)+\int_{-\infty}^xdx'\Pi(x')\right]\right\}\,.
\end{equation} 
Here $v=v_F/g$ is the electron velocity renormalized by the interaction, $p_F=\mu_{T\!I}/v_F$, $\hbar \omega_c$ is the energy cutoff. 

Electrons in the strip are supposed to be non-interacting gas described by the Hamiltonian
\begin{equation}
H_N=\sum_{k,n}\left(\frac{\hbar^2k^2}{2m}+\varepsilon_n+E_c\right)a^{\dag}_{k,n}a_{k,n}\,,
\end{equation} 
where $\varepsilon_n$ is the quantization energy of $n$th subband, $a^{\dag}_{k,n}$ and $a_{k,n}$ are creation and annihilation operators, $E_c$ is the energy of the conduction band bottom in the infinite 2D electron gas.

The tunneling Hamiltonian is conveniently written in momentum presentation, i.e. in terms of the operators $a^{\dag}_{k,n}, a_{k,n}$ and Fourier harmonics of the field operators $\psi^{\dag}_{1D}, \psi_{1D}$~\cite{Governale1}
\begin{equation}
 H_{tunn}=\sum_{p,k,n}T_n(p,k)c^{\dag}_pa_{k,n}+H.c.\,,
\end{equation} 
with $c^{\dag}_p$ being the creation operator for 1D fermion with momentum $p$. The tunneling matrix element is easy to find in the case where the strip is a square potential well of the width larger than the penetration length of electrons under the barrier,
\begin{equation}
 |T_n(p,k)|^2\approx\mathcal T^2\frac{\pi^2 n^2 L^2}{l_Nl_{ES}W^3}\left(\frac{\sin[(p-k)L/2]}{(p-k)L/2}\right)^2,
\end{equation} 
with $l_N$ and $l_{ES}$ being the normalization lengths for the strip and edge states.

The electron and spin currents in the system are compactly expressed in terms of a matrix $I^{r,l}$ defined as a tunnel flow of electrons from chiral state ($r=\pm$) in the TI to the right- and left-moving states ($l=\pm$) in the strip. We calculate $I^{r,l}$ in the first approximation in $H_{tunn}$. The standard procedure~\cite{Mahan} gives an expression for $I^{r,l}$ in terms of spectral functions and the Fermi distribution function $f(\epsilon)=1/[1+\exp(\epsilon/T)]$:
\begin{equation}
\label{chiral_current}
\begin{array}{rl}
I^{r,l}= &\frac{l}{\hbar}\sum \limits_{l k>0}\sum \limits_{p,n}|T_n(p,k)|^2\!\int\!\frac{d\epsilon}{2\pi}A_n(k,\epsilon+\!eV)\\
&\times A_{ES}^r(p,\epsilon)[f(\epsilon)\!-\!f(\epsilon\!+\!eV)]\,.
\end{array}
\end{equation}
Here $A_n(k,\epsilon)$ is spectral functions of non-interacting electrons in the strip
\begin{equation}
 A_n(k,\epsilon)=2\pi \delta\left(\epsilon-\frac{\hbar^2k^2}{2m}-\varepsilon_n-E_c+\mu_{T\!I}\right)\,;
\end{equation}
$A_{ES}^r(p,\epsilon)$ is the spectral function of the chiral 1D fermions~\cite{Governale1}
\begin{equation}
\begin{array}{rl}
 A_{ES}^r(p,\epsilon)=&\!\!\frac{\hbar \omega_c}{4\pi}\left[K_{\alpha}\!\left(\!\frac{\epsilon-r\hbar
v(p-rp_F)}{2}\!\right) K_{\alpha+1}\!\left(\!\frac{\epsilon+r\hbar v(p-rp_F)}{2}\!\right)\right.\\
& \biggl.+(\epsilon\to -\epsilon,\, p-rp_F\to -p+rp_F)\biggr]\,,
\end{array}
\end{equation}  
where $\alpha=(g+1/g-2)/4$, 
\begin{equation}
K_{\alpha}(\epsilon)=\frac{e^{\epsilon /2T}}{\hbar \omega_c\Gamma(\alpha)}
\left(\frac{\hbar \omega_c}{2\pi T}\right)^{1-\alpha}
\left|\Gamma \left(\frac{\alpha}{2}+\frac{i\epsilon}{2\pi T} \right)\right|^2\,,
\end{equation} 
with $\Gamma(\alpha)$ being the gamma function.

The tunnel current $J$ through the contact is given by
\begin{equation}
 J=-e\sum_{r=\pm}\sum_{l=\pm} l I^{r,l}\,.
\end{equation} 
The electron currents $J^l$ in the right ($l=+1$) and left ($l=-1$) branches of the strip outside the contact are determined as follows
\begin{equation}
J^l= -e \sum_{r=\pm}I^{r,l}\,.
\end{equation} 
The spin current $J_S^{l}$ in the right and left branches of the strip reads
\begin{equation}
 J_S^l=\frac{\hbar}{2}\sum_{r=\pm}rI^{r,l}\,.
\end{equation} 
The spin polarization of the current in the strip is defined as the ratio of the spin over the charge current in each branch
\begin{equation}
 P_l=\frac{2e}{\hbar}\frac{J_S^l}{J^l}
\end{equation} 

From the symmetry of the matrix $I^{r,l}$ [see Eq.~(\ref{chiral_current})] it is evident that $P_{-l}=-P_l$.

\section{Results and discussion}\label{results}
We have calculated the electron and spin currents and studied in detail their dependence on such parameters as the voltage drop on the tunnel contact, the size of the contact, the interaction parameter in the edge states and the chemical potential of the system.

\begin{figure}
 \centerline{\includegraphics[width=8cm]{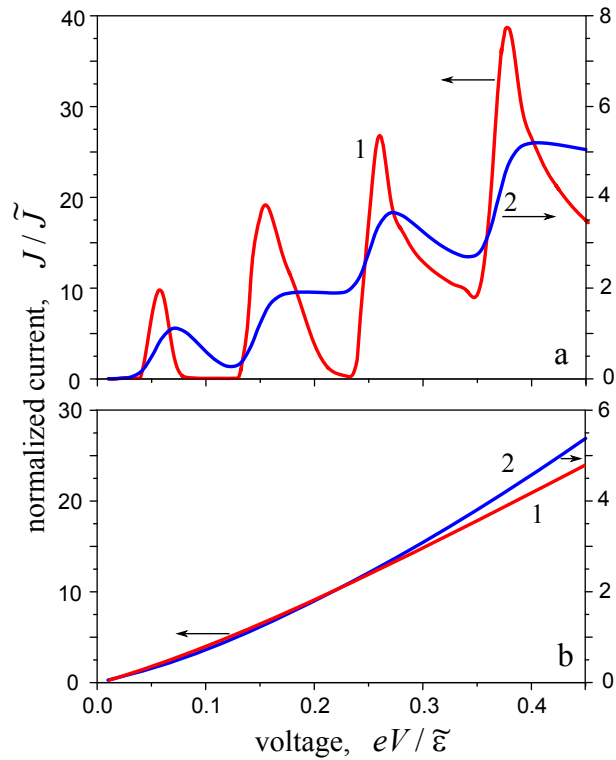}}
 \caption{(Color online) Tunnel current   through the contact as a function of the voltage for different strip width $W$ and the interaction parameter $g$. (a) $W=30 \tilde{l}$, $g=0.9$ (line 1), $g=0.6$ (line 2). (b) $W=500 \tilde{l}$, $g=0.9$ (line 1), $g=0.6$ (line 2). Other parameters are: $E_c=0$, $\mu=0.5\tilde{\varepsilon}$, $T= 0.01 \tilde{\varepsilon}$,  $L=300 \tilde{l}$.}
\label{J-V}
\end{figure}

The electron current $J$ through the contact as a function of the applied voltage $V$ is shown in Fig.~\ref{J-V} for small voltage ($eV <\mu_{TI}$). In what follows, the dimensionless notations are used: the current, the spin current, the length and the energy are normalized correspondingly to $\tilde{J}=2em^3 v_F^2\mathcal T^2L/\pi \hbar^5$, $\tilde{J}_s=\tilde{J}\hbar/2e$, $\tilde l=\hbar/2mv_F$, $\tilde\varepsilon=2mv_F^2$. The voltage is normalized to $\tilde{\varepsilon}/e$.

The results presented in Fig.~\ref{J-V} were obtained in the case of long contact $Lp_F\gg 1$ ($p_F$ being the Fermi wavevector in the edge state) where the scattering of the momentum is small and therefore the polarization is close to unity. The current oscillations in Fig.~\ref{J-V}a are caused by the quantization on the strip width. Increasing the interaction strength shifts the first peak to higher voltage. The peaks of the current are smoothed out with increasing temperature and decreasing $g$ and $L$.

In the case of the wide strip, $W\gg W_c=4\pi\hbar/\sqrt{m\cdot\max(eV,T)}$, the oscillations disappear (Fig.~\ref{J-V}b) and the current-voltage characteristic turns into a power dependence (Fig.~\ref{J-V}b) that is standard for tunneling in the Luttinger liquid~\cite{Gimarchi}.

A similar oscillatory behavior occurs also in the dependence of the current on the chemical potential.

\begin{figure}
 \centerline{\includegraphics[width=8cm]{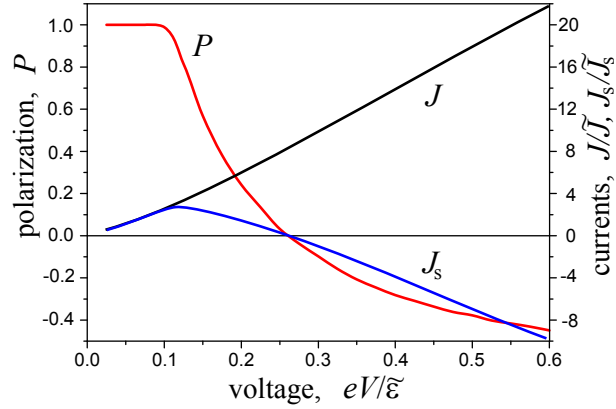}}
 \caption{(Color online) Spin polarization, the charge and spin currents in the left branch of the strip as a function of the voltage. The used parameters are: $W=200 \tilde{l}$, $L=500 \tilde{l}$, $E_c=0$, $\mu_{T\!I}=0.1\tilde{\varepsilon}$, $T=0.01\tilde{\varepsilon}$, $g=0.9$.}
\label{P-V}
\end{figure}

The most interesting result is obtained in calculations of the spin polarization of the current in the strip. Fig.~\ref{P-V} shows the spin and electron currents as a function of voltage for a rather wide strip ($W\gg W_c$). At small voltage, the spin current practically coincides with the electron current, however with increasing voltage the spin current drops to zero and then changes its sign. Correspondingly, the polarization changes from unity to a negative value, which goes asymptotically to -1. This behavior of $P$ is explained as follows. When the voltage is small, the energy layer $eV$, within which current-carrying electrons tunnel from the TI into the strip, lies above the Dirac point. In this case, practically all electrons with spin down come into left-moving states and spin-up electrons come to right-moving states in the strip (see Fig.~\ref{TI-N_contact}b) and therefore $|P|\approx1$. With increasing voltage, the above layer of tunneling electrons expands downward. When its lower boundary intersects the Dirac point, a fraction of the tunneling electrons with spin up comes into the left-moving states (see Fig.~\ref{TI-N_contact}c) and similarly some spin-down electrons come to the right-moving states. As a result, the polarization decreases and changes the sign.

A similar effect takes place also in the case of the polarization dependence on the chemical potential. The polarization changes the sign when $\mu_{T\!I}$ is decreased below the Dirac point. The width of the transition interval increases with increasing $V$ and $T$ and with decreasing contact length $L$.

\begin{figure}
 \centerline{\includegraphics[width=8cm]{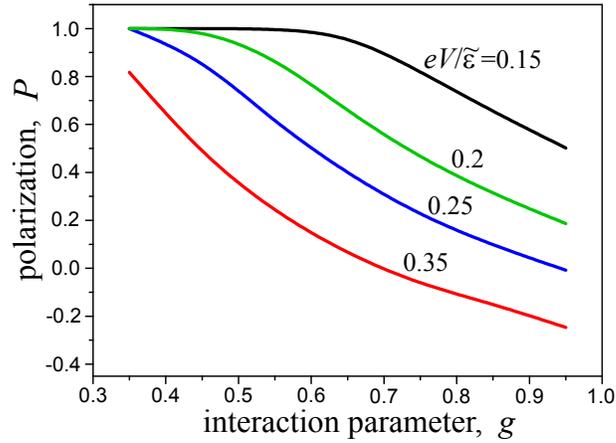}}
 \caption{(Color online) The dependence of the polarization on the electron-electron interaction parameter for various voltages. The used parameters are: $W=200\tilde{l}$, $L=500\tilde{l}$, $E_c=0$, $\mu_{T\!I}=0.1\tilde{\varepsilon}$, $T=0.01\tilde{\varepsilon}$, $g=0.9$.}
\label{P-g-V}
\end{figure}

An interesting question is how the electron-electron interaction strength affects the polarization. The interaction effect is described by the parameter $g$, which decreases from 1 to 0 on increasing the interaction potential from zero to infinity. Within the Luttinger liquid model, the decrease in $g$ has two consequences that could be important for the current polarization. First, the velocity of the chiral fermions increases and second, the 1D fermions are distributed over wider region of momentum space.

The dependence of $P$ on $g$ is presented in Fig.~\ref{P-g-V}. The polarization is seen to increase with decreasing $g$. This fact is explained by the first of above factors, namely, by an increase in the velocity. Actually, increasing the velocity leads to a shift of the Dirac point down and therefore a fewer electrons fall into the energy interval below the Dirac point. Fig.~\ref{P-g-V} shows also that at high enough voltage the polarization sign changes with increasing $g$.
 
\begin{figure}
 \centerline{\includegraphics[width=8cm]{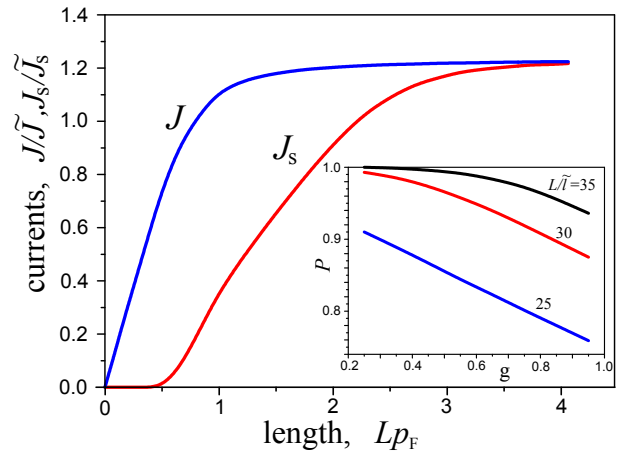}}
 \caption{(Color online) The electron and spin currents as a function of the contact length, $W=150 \tilde{l}$, $\mu_{T\!I}=0.12\tilde{\varepsilon}$, $E_c=0$, $T=0.01\tilde{\varepsilon}$, $g=0.9$, $eV=0.05\tilde{\varepsilon}$. Inset: the dependence of the polarization on the interaction parameter for various lengths $L$.}
\label{J-L-g}
\end{figure}

Now consider the dependence of the polarization on the length of the contact, which reflects the effect of the electron backscattering by the contact of small length. Fig.~\ref{J-L-g} shows the dependence of the charge and spin currents in the strip on the contact length. As the contact length is small ($Lp_F\ll 1$) the charge current grows with $L$ as $J\sim L^2$~\cite{Note}  and the spin current is negligibly small because of strong scattering by the contact. When $Lp_F\gtrsim 1$ the spin current increases, because the scattering becomes weaker, and asymptotically reaches the electron current at $Lk_F\gg 1$, with $k_F=\sqrt{2m(\mu_{T\!I}-E_c-\varepsilon_1)}/\hbar$. In this limit, the currents $J$ and $J_S$ increase linearly with $L$ until the perturbation theory in the tunnel Hamiltonian is not disturbed~\cite{Governale2}. The inset shows that the polarization decreases with decreasing  interaction strength similarly to Fig.~\ref{P-g-V}.

\section{Conclusion}\label{conclude}
We have studied the electron tunneling from helical edge states of a 2D TI to a normal 2D electron gas and concluded that this process can be an effective mechanism of creating the spin current in normal conductors.

The mechanism consists in the fact that the electron tunneling into the normal conductor gives rise to two oppositely directed currents of electrons with opposite spins. One current is directed to the right from the contact, and the other does to the left, so that an single continuous spin current flows parallel to the common boundary of the TI and the normal electron gas.

We have studied in detail this mechanism in detail for the system where a 2D strip of normal electron gas is coupled to the 2D TI via a tunnel contact of finite length. The electron current along the strip is equal half of the tunnel current through the contact and is highly polarized. The spin polarization essentially depends on the contact length, the electron-electron interaction parameter of the edge states and the voltage drop across the contact.

The most significant features of the spin-polarization behavior are as follows:
\begin{enumerate}
 \item the polarization increases with increasing contact length on the scales $Lp_F \sim 1$ or $Lk_F \sim 1$;
 \item the direction of the spin current changes with changing voltage or the chemical potential when the bottom of the subbands in the strip crosses the Dirac point of the edge state spectrum;
 \item the polarization grows with increasing electron-electron interaction strength due to the renormalization of the electron velocity caused by the interaction.
\end{enumerate}

The effect we have described here could be observed experimentally, for example, by measuring the spin accumulation near contacts to the strip.  So, if the strip is connected with electron reservoirs containing localized spin states, which are a spin bath, the spin density is accumulated there, in the same way as has been shown recently in Ref.~\cite{Lunde}.

\section*{Acknowledgments}
This work was supported by the Russian Foundation for Basic Research (project No 11-02-00337) and Programs of the Russian Academy
of Sciences.


\section*{References}
\begin{thebibliography}{20}
\bibitem{Hasan}Hasan M Z and Kane C L 2010 Rev. Mod. Phys. \textbf{82} 3045 
\bibitem{Qi}Qi X L and Zhang S C 2011 Rev. Mod. Phys. \textbf{83} 1057 
\bibitem{Konig}K\"onig M, Wiedmann S, Br\"une C, Roth A, Buhmann H, Molenkamp L W, Qi X-L and Zhang Sh-Ch 2007 Science \textbf{318} 766
\bibitem{Roth}Roth A, Bruene C, Buhmann H, Molenkamp L W, Maciejko J, Qi X-L and Zhang Sh-Ch 2009
Science \textbf{325} 294 
\bibitem{Brune}Br\"une C, Roth A, Buhmann H, Hankiewicz E M, Molenkamp L W, Maciejko J, Qi X-L and Zhang Sh-Ch 2012 Nature Physics \textbf{8} 486 
\bibitem{Gusev}Gusev G M, Olshanetsky E B, Kvon Z D, Levin A D, Mikhailov N N and Dvoretsky S A 2012 Phys. Rev. Lett. \textbf{108} 226804 
\bibitem{Kong}Kong D, Jason C Randel J C, Peng H, Cha J J, Meister S, Lai K, Chen Y, Shen Zh-X, Manoharan H C and Cui Y 2010  Nano Lett. \textbf{10} 329
\bibitem{Wu}Wu C, Bernevig B A and Zhang Sh-Ch 2006 Phys. Rev. Lett. \textbf{96} 106401
\bibitem{Gimarchi}Giamarchi T 2004 \textit{Quantum Physics in One Dimension} (Oxford University Press, Oxford)
\bibitem{Auslander}See, e.g., Auslaender O M, Yacoby A, de Picciotto R, Baldwin K W, Pfeiffer L N and West K W 2002 Science \textbf{295} 825
\bibitem{Governale1}Governale M, Grifoni M and Schon G 2000  Phys. Rev. B \textbf{62} 15996
\bibitem{Boese}Boese D, Governale M, Rosch A and Z\"ulicke U 2001 Phys. Rev. B 64 085315 
\bibitem{Carpentier}Carpentier D, Pe\c{c}a C and Balents L 2002  Phys. Rev. B \textbf{66} 153304 
\bibitem{Governale2}Governale M, Boese D, Z\"ulicke U and Schroll C 2002 Phys. Rev. B \textbf{65} 140403(R) 
\bibitem{Dolcetto}Dolcetto G, Barbarino S, Ferraro D, Magnoli N  and Sassetti M 2012 Phys. Rev. B \textbf{85} 195138 
\bibitem{SST}Sukhanov A, Sablikov V and Tkach Yu 2009 J. Phys. C \textbf{21} 375801
\bibitem{Voit}Voit J 1995 Rep. Prog. Phys. \textbf{58} 977 
\bibitem{Mahan}Mahan G D 2000 \textit{Many Particle Physics} (3 rd edition, Kluwer/Plenum)
\bibitem{Note}Note that the current $J$ is normalized to $\tilde J$, which is proportional to $L$
\bibitem{Lunde}Lunde A M and Platero G 2012  Phys. Rev. B \textbf{86} 035112
\end{thebibliography}
\end{document}